# Charged Particle Detection using a CMOS Active Pixel Sensor

Howard S. Matis, Fred Bieser, Stuart Kleinfelder, *Member, IEEE*, Gulshan Rai, Fabrice Retiere, Hans Georg Ritter, Kunal Singh, Samuel E. Wurzel, Howard Wieman, and Eugene Yamamoto

*Abstract*--Active Pixel Sensor (APS) technology has shown promise for next-generation vertex detectors. This paper discusses the design and testing of two generations of APS chips. Both are arrays of 128 by 128 pixels, each 20 by 20 μm. Each array is divided into sub-arrays in which different sensor structures (4 in the first version and 16 in the second) and/or readout circuits are employed. Measurements of several of these structures under $Fe^{55}$ exposure are reported. The sensors have also been irradiated by 55 MeV protons to test for radiation damage. The radiation increased the noise and reduced the signal. The noise can be explained by shot noise from the increased leakage current and the reduction in signal is due to charge being trapped in the epi layer. Nevertheless, the radiation effect is small for the expected exposures at RHIC and RHIC II. Finally, we describe our concept for mechanically supporting a thin silicon wafer in an actual detector.

## I. INTRODUCTION

MODERN collider detectors frequently need to measure a vertex that has an origin away from the collision point. Vertex detectors provide tracking information to decide whether a track comes from the primary vertex or from a secondary decay [1]. With impact resolution in the tens of microns, they can identify particles with cτ of 100's of microns. Consequently, they are ideal to detect mesons with charm or bottom quarks, which have these decay properties.

For example, the SLD collaboration built a vertex detector [2] based on CCD technology [3]. Such pixel detectors have the advantage of simultaneously measuring all three space point coordinates. They do not have the hit ambiguity problem of drift detectors. The small pixel size provides excellent spatial resolution. Placing the detector as close to the beam collision point also improves the resolution, because it reduces the track extrapolation error. As the detector is in a low radiation environment, CCDs could be used at this accelerator. CCDs require that the charge be transferred from one pixel to another. Charge in the end row of a pixel chip, for example a 1000 × 1000 array, must be transferred through more than 1000 pixels before being digitized. Therefore, any small loss in charge transfer produces large signal loss and signal sharing.

Because of the high radiation environment and need to have the vertex detector in the trigger, CCDs are not the appropriate choice [4] at LHC (Large Hadron Collider) at CERN. At the LHC, the three major experiments [5] decided to use a hybrid technology where the sensor is bump bonded to a read-out chip. The hybrid technology has the disadvantage that the pixel size is much greater than a CCD pixel and that two chips have to be assembled. The two chips and their interconnection are much thicker than can be done in CCD technology.

Through research by the LEPSI/IReS group [6], Active Pixel Sensors [7] have recently emerged as a competitor to CCDs and the hybrid technology for charged-particle pixel detectors. Like CCDs, APS detectors can be built with thin wafers and with small pixels. Unlike CCDs, charge is directly read out from each pixel without shifting through the rest of the detector. In principle, APS detectors can operate in much higher radiation environments than CCDs. Furthermore, as they can be built in standard CMOS, features such as ADCs and zero suppression can be put in the periphery of the chip. For example, to make a high-speed APS sensor, [8] put an ADC on each pixel.

To produce an APS detector for charged particles, an epitaxial silicon (epi) layer is used as a deep charge collection region. When a charged particle traverses the APS sensor, it creates electron-hole pairs in the epi layer. As the epi region can be much thicker than a conventional APS diode, a greater amount of charge can be liberated and collected. However, as the epi layer is field free, the holes diffuse until they reach the $p^+$ bulk region, while the electrons diffuse until they reach a pixel's $n^+$ diode. Because of this phenomenon, hits spread out over several pixels, while CCDs tend to collect the charge in one or two pixels.

## II. THE STAR DETECTOR

The STAR Collaboration [9] is examining whether APS technology is appropriate for an inner vertex detector [10]. That detector is currently running at RHIC (Relativistic Heavy Ion Collider), which is operated by Brookhaven National Laboratory. The focus of the detector is to study collisions between circulating Au beams at 100 GeV·A. Initial

Manuscript received December 2, 2002. This work was supported in part by the Director, Office of Science, of the U.S. Department of Energy under Contract No. DE-AC03-76SF00098.

H. S. Matis is with the Lawrence Berkeley National Laboratory, Berkeley, CA 94720 USA (telephone: 510-486-5031, e-mail: hsmatis@lbl.gov).

F. Bieser, G. Rai, F. Retiere, H. G. Ritter, K. Singh. S. E. Wurzel, H. Wieman, E. Yamamoto are with the Lawrence Berkeley National Laboratory, Berkeley, CA 94720 USA.

S. Kleinfelder is with the Department of Electrical and Computer Engineering, University of California, Irvine, CA 96297 USA.

measurements at this energy have been completed. As there have been recent technical progress with vertex detectors, it is now conceivable that detailed measurements on charmed quarks can be made.

Current theoretical work indicates that measurements of charmed quarks are very appealing [11] as they are produced almost exclusively from initial state parton-parton interactions, while lighter quarks (u, d, s) may be produced in hadronic interactions. By this virtue, charm provides a more direct connection to the early stage without contamination from later phases. The total charm yield should be sensitive to the initial state, while the hadronic charm composition ($J/\psi$, D, $D_s$, $\Lambda_c$) will depend on the dynamical evolution of the system. These particles can be a signature of the Quark Gluon Plasma. Since their mass is heavy, it is much easier to calculate their production.

Simulations show that the STAR detector could detect charmed particles but not produce differential cross sections. The STAR detector does contain a vertex detector called the SVT (Silicon Vertex Tracker). The SVT uses three layers of silicon drift detectors to measure the position of a track. These detectors are relatively thick (a few percent of a radiation length), far (> 6 cm) from the interaction point, and have a predicted position resolution of 20 µm. To study the physics of charm, a high-resolution inner vertex detector is needed in STAR.

We have been simulating a hypothetical vertex detector with thickness of 80 µm and resolution of 4 µm. There are two cylindrical detectors at radii of 2.8 cm and 3.82 cm away from the interaction point. Inside the detector, there is Be beam pipe with a radius of 2.2 cm. Simulations show that with such a detector an invariant cross section from the $D^0$ meson can be measured. What follows in this paper is our work to investigate whether an APS detector is appropriate for accelerator experiments.

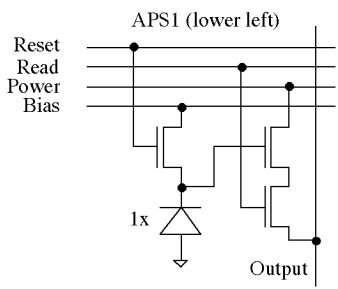

Fig. 1. The circuit diagram shows an APS pixel circuit that was tested in this paper.

### III. CHIP CONFIGURATIONS

Two CMOS radiation sensor ICs, APS-1 and APS-2, have been designed, fabricated, and tested. Each prototype sensor array includes 128 by 128 pixels with a pixel size of 20 by 20 µm. Each array is about 2.5 mm on a side. Both chips were designed in a standard TSMC digital 0.25 µm CMOS process that includes an 8-10 µm epitaxial layer. The layouts of the chip and some previous results with a 1.5 GeV electron beam have been discussed in [12].

In this paper, we will discuss APS-2, which has 16 test structures. We will concentrate on four standard APS configurations, one of which is shown in Fig. 1. The structures have 1, 2, 3 or 4 pickup diodes. In general, we have found similar results when comparing APS-1 to APS-2.

### IV. TESTS WITH $FE^{55}$

To record the data from the APS chip, we built a test DAQ board to digitize and store the data. The output of the APS went to an ADC, which digitized the data at 0.4 MHz into 16 bits. All of the data presented below is taken at room temperature. We use the correlated double sample method to remove and reduce fixed pattern reset noise by subtracting subsequent frames. As the chip is not reset in between reads, the difference is simply the integrated charge in the diode, and reset noise is canceled. Fig. 2 shows a typical spectrum from $Fe^{55}$. To create this histogram, we use a very simple algorithm that looks for the highest ADC value and then sum over a square array of pixels, for example 5 × 5 pixels, such that the peak pixel is at the center of that array. After making that sum, we zero those pixels and then repeat the procedure. We stop looking for hits when the highest pixel is less than a pre-determined threshold. A more sophisticated algorithm could produce results, which might enhance the performance on the detector.

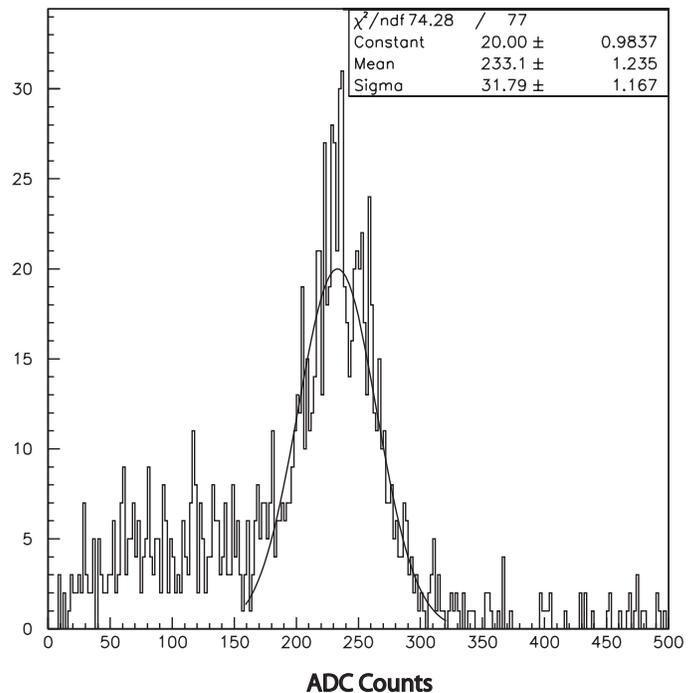

Fig. 2. This is a histogram of a typical ADC spectrum for summing 5 × 5 pixels. The curve shows the fit to the 5.9 keV x-ray line. For this plot, the threshold for calculating a sum is 20 ADC counts.

Similarly, we also do sums of 3 × 3 (9 pixels) and 7 × 7 (49 pixels). To sum 4 pixels, we take the 3 × 3 array and then find the highest 2 × 2 sum that contains the center pixel. We take the highest 4-pixel array and then find the highest 3 pixels to find the 3-pixel sum. We use a similar method to find the 2-pixel sum. Monte Carlo studies have shown that the 2, 3, and 4 pixels sums are biased because they are susceptible to noise fluctuations. As the noise is comparable to the charge collected in the outside pixels, the algorithm tends to pick those pixels where the noise is larger. Consequently, the 9 pixel sums are a more accurate measurement of the energy of an event than the lower pixel sums. The counts with higher ADC values are produced by pileup events. The reconstruction algorithm is very simple and does not reject those events.

The diffusion of the electrons in the epi-layer can be studied by comparing different numbers of pixel sums. Fig. 3 shows the collected charge for various sums. The single pixel sum clearly shows a peak at 5.9 keV. Higher statistical studies show the less frequent 6.5 keV line. These peaks are produced when the γ-ray converts near the $n^+$ diode and all the charge is collected. If the x-ray does not convert near the diode and coverts in the epi layer, the charge diffuses in the epi layer. The various pixel sums show the extent of diffusion. This data show that a 5 × 5 array captures most of the charge, but not all.

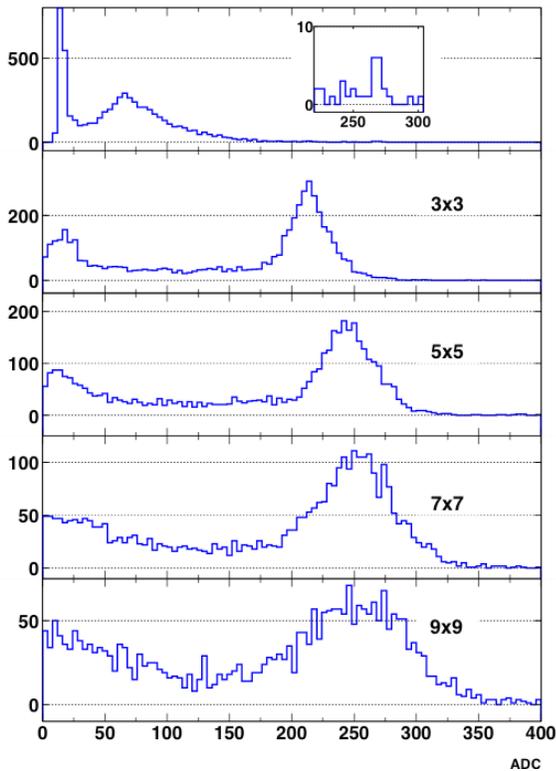

Fig. 3. Various $Fe^{55}$ spectra for different pixel sums. The top curve is for a single pixel. The other curves represent a square pixel array centered on the highest pixel. Each graph has a label that indicates the square pixel sum area. The insert in the top graph shows the 5.9 keV peak at a higher scale. The x-axis shows the number of ADC counts. The $Fe^{55}$ peak can be seen in each sum.

We define signal to noise, as the mean charge in the $Fe^{55}$ peak divided by $\sigma_1\sqrt{n}$, where $n$ is the number of pixels summed and $\sigma_1$ is the sigma of the noise for a single pixel. Fig. 4 shows this ratio for different number of diodes attached to the standard APS configuration. From this data, we conclude that the single diode structure has slightly better over-all signal to noise ratio then the other configurations. It is apparent that the extra charge collected by the diodes has less of an effect than the increased capacitance of the diodes. The data presented in this figure have the one diode near the other transistors of the APS circuit. We found that if we centered the diode in the middle of the pixel, the signal to noise ratio is worse. This reduction occurs because the extra capacitance of the longer trace reduces the collected charge.

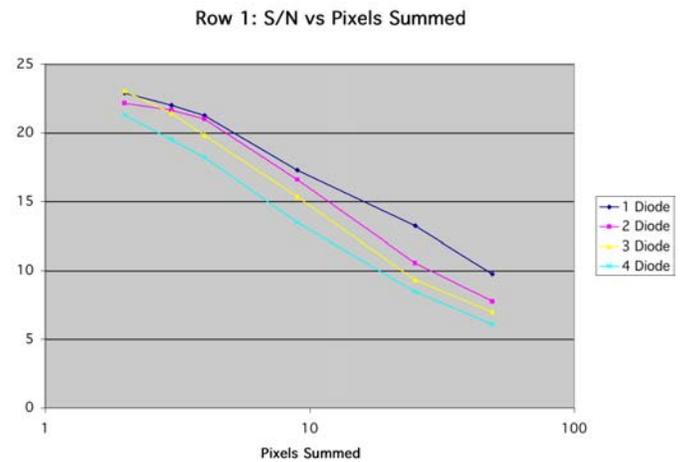

Fig. 4. Signal to Noise for different diode configurations.

## V. RADIATION EFFECTS

To determine the effect of radiation, we exposed the chips to 55 MeV protons at the Lawrence Berkeley Laboratory 88" Cyclotron. Each APS chip was mounted in a chip carrier. The center of the chip was about 2.38 cm away from the beam center. The intensity of the beam was monitored by the standard 88" beam diagnostics in beam line 3B. The diagnostic program measures the fluence in protons/cm². To scale from the low energy proton exposure to that of RHIC, we used the NIEL scaling hypothesis that is described in [13].

Table 1 shows the exposures for the various chips. We use the conversion that 1 rad = $6.7 \times 10^6$ protons/cm². We assume that a RHIC year provides collisions for a continuous total of 20 weeks and that RHIC II has a luminosity 40 times RHIC.

We measured the leakage currents before and after the radiation exposure. The leakage current before the exposure was approximately the same for each chip.

Whenever we measure a hit, we always subtract the mean leakage current for each pixel. This assumes that the leakage current does not vary. Consequently, any fluctuation in the leakage current will contribute to the increased noise.

Therefore, this variation is what contributes to detector performance. The time to read each pixel was 2.56 µs and the total time measured for the leakage current was 2.63 ms. The maximum speed of the APS chip is about 1 Mpixel/s, which, if used, would reduce measured leakage. Leakage current can be corrected. However, if the leakage current were high, then the charge variation (shot noise) would increase the measured noise. Furthermore, the larger the leakage current, the chip needs to reset more frequently so that the dynamic range is not exceeded.

TABLE 1
RADIATION EXPOSURE FOR TEST AT THE 88" CYCLOTRON. RHIC EXPOSURE IS THE EQUIVALENT NUMBER OF YEARS ASSUMING NOMINAL OPERATING CONDITIONS. RHIC II IS THE PROJECTED RADIATION DOSE FOR THE NEW MACHINE THAT IS 40 TIMES THE LUMINOSITY OF RHIC.

| Exposure | Proton Flux ($\times 10^{12}$ cm$^2$) | Equivalent Dose (krad) | RHIC Exposure (y) | RHIC II Exposure (y) |
|---|---|---|---|---|
| 1 | 0.144 | 21 | 18 | 0.5 |
| 2 | 0.485 | 72 | 62 | 1.5 |
| 3 | 0.96 | 143 | 122 | 3.1 |
| 4 | 3.02 | 451 | 385 | 9.6 |
| 5 | 9.88 | 1475 | 1259 | 31.5 |

Once we correct for the leakage current, we can then see the radiation exposure's effect on signal and noise on the performance of the chips. Fig. 5 shows the results. To determine the $Fe^{55}$ peak, we use the same technique as previously described. We were able to measure a clear peak for all exposures except the highest at $1.0 \times 10^{13}$ p/cm$^2$. The data show a decrease in pulse height and a gradual increase of noise.

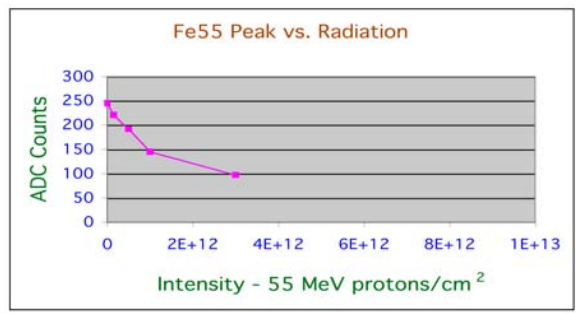

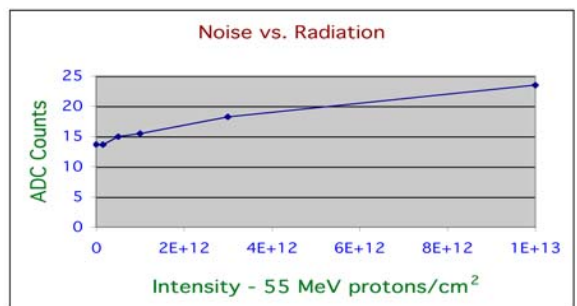

Fig. 5. The top graph shows the $Fe^{55}$ signal as a function of fluence, while the bottom graph shows the increase in noise.

Radiation induced bulk damage in the epi layer can explain some of the loss of signal. The traps can capture the diffusing electrons and prevent them from being collected by the APS diode. To determine where the charge is lost, we measured the response to $Fe^{55}$ of an irradiated detector. Figs. 6a-b show the results of an unexposed detector, while 6c-d show one that exposed to 143 krad. Both Figs. 6a and 6c demonstrate that the 5.9 keV x-ray peak occurs at the same place. Therefore, the basic CMOS operation is not compromised and the gain in the diode is not affected. However, Figs. 6b and 6d show there is a shift in the $5 \times 5$ pixel sum for the irradiated chip. This shift occurs in the charge collected from the epi layer, and therefore it implies that charge is lost in the epi layer. The LEPSI/IReS group has made similar measurements [14] with neutron radiation. Their conclusions for the effect of radiation damage are consistent with ours.

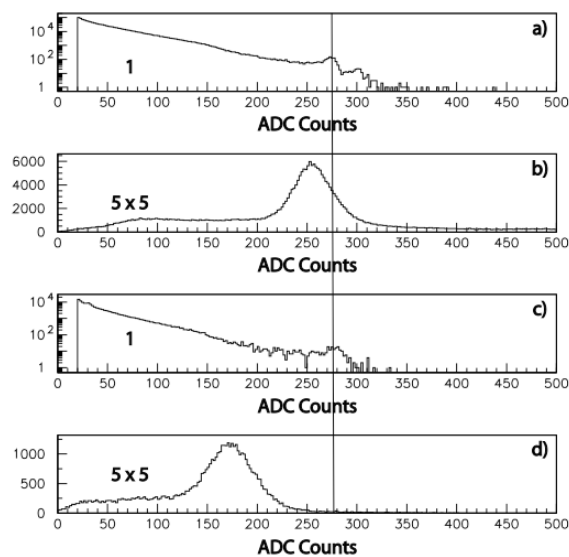

Fig. 6. Comparison of a detector that was exposed to 143 krad of 55 MeV protons to an unexposed detector. Graph a) shows the single hit charge collected for an unexposed chip, while b) shows the sum for a $5 \times 5$ array. Similarly, c) shows the single hit for the irradiated detector and d) the 5 x 5 sum. The vertical line shows the location of the $Fe^{55}$ 5.9 keV x-ray, The peak position of the 5.9 keV peak is in the same location for a) and c), while the charge collected through the epi-layer, as shown in the $5 \times 5$ sum is different. The higher energy 6.5 keV $Fe^{55}$ line can be seen in a) and not c) because a) has an order of magnitude more events. These plots were taken at a higher rate than the other $Fe^{55}$ data, so there is a more significant pileup effect.

To study the source of the noise, we calculated the contribution caused by the shot noise of the leakage current. We then subtract the shot noise and look at difference. These results are shown in Fig. 7. As the difference is roughly constant with leakage current or radiation fluence, the increase in noise can be attributed mostly to shot noise. As the leakage current decreases with readout speed, reading the chips faster would result in less noise. Once again, the extra leakage current is only important, if the shot noise exceeds the other contributions. In this figure, we have converted the scale into electrons. To do this, we assume all of the charge of the $Fe^{55}$ is collected by in the 5.9 keV peak. This corresponds to 1638 electrons.

It is clear that this increase in noise and reduction of signal might restrict the use of APS technology. To explore its use in a potential accelerator environment, we examine the impact of its use at RHIC. Exposure #1 is the estimated equivalent of 18 years at RHIC, while the exposure #2 is projected to be equivalent to 1.5 years at RHIC II. These data show that the reduction on performance of the chip is relatively small. As mechanical supports could be designed so that the silicon can be replaced each year, the detector need only last one year in an accelerator environment until a convenient accelerator maintenance period occurs. From these tests, we have a strong indication that APS technology can withstand the radiation environment of RHIC.

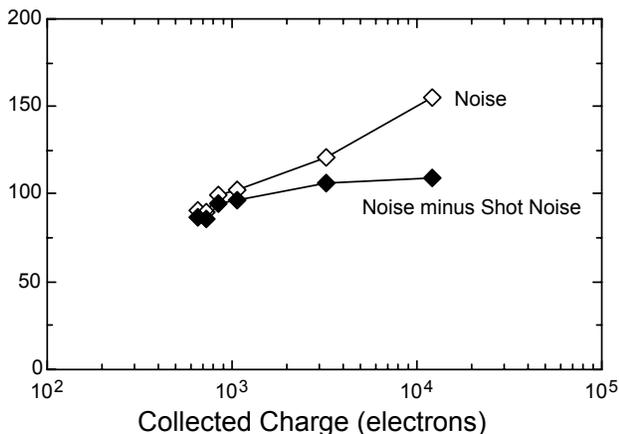

Fig. 7. This upper line shows the variation of noise with leakage current. The lower line is the result when the shot noise is subtracted. The y-axis is in units of electrons.

## VI. MECHANICAL DESIGN

Because we want to minimize the mass of the detector, we have had some sample wafers thinned to 50 μm and are currently developing methods for handling and supporting the thinned silicon strips under tension. As the $p^+$ substrate mostly provides mechanical support for the device, it is possible to remove it and have the same sensitivity to charged particles. In fact, it is common, in astrophysical CCD applications, to remove this substrate [15] and sometimes even remove part of the epi layer so that the back of the chip can be illuminated.

Our concept for a detector support is illustrated in Fig. 8. In this design 10 cm silicon detector strips or ladders and aluminum Kapton flex cables are supported under tension by the gray structures at either end. The ladders, consisting of five 1.6 × 1.6 cm CMOS chips, are shown in blue (darker shading). The flex cables are shown in yellow (lighter shading). As shown in the left side of the figure, there are two detection layers, one at an inner radius and one at an outer radius. The 24 ladders are arranged in modules of 3 ladders as shown on the right side of the figure. The detector unit is supported at one end only so that the whole assembly can be easily removed and replaced should the primary beam stray.

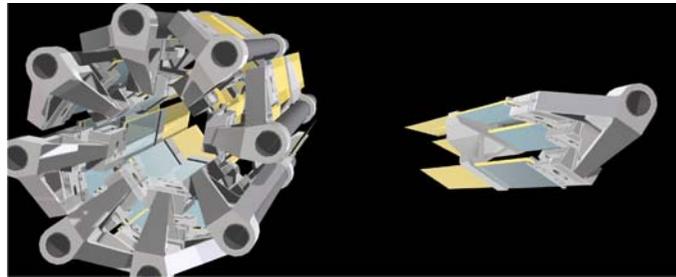

Fig. 8. The right picture shows the mechanical concept for mounting three ladders of silicon. The blue area (darker shaded area) represents the silicon, while the yellow region (lighter shaded region) that extends past the silicon is the aluminum-Kapton cable.

## VII. SUMMARY

Our results show that APS technology is very promising for developing a vertex detector. Our chips can detect particles from x-rays to electrons. Unlike CCDs, charge for APS chips diffuse to several pixels. Consequently, the intrinsic signal to noise is less for APS chips, as many pixels need to be summed as charge diffuses in the epi layer. Radiation tests show that the APS technology should be radiation resistant under nominal RHIC operating conditions. When RHIC II becomes operational, there would be only a small decrease in signal and increase in noise over a three-year exposure. Mechanical prototypes are under construction and will soon be studied to ascertain a practical method of supporting very thin silicon.

## VIII. ACKNOWLEDGMENT

We would like to thank Peggy McMahan and the staff of Lawrence Berkeley National Laboratory's 88" cyclotron for their assistance for the radiation exposure. John Wolf assembled the data acquisition board and several of our test fixtures.